\documentclass[a4paper,11pt,onecolumn]{article}
\usepackage{graphicx}
\usepackage{dcolumn}
\usepackage{bm}
\usepackage{psfig}
\def\be{\begin{equation}}
\def\ee{\end{equation}}
\def\beq{\begin{eqnarray}}
\def\eeq{\end{eqnarray}}
\def\bn{\begin{eqnarray*}}
\def\en{\end{eqnarray*}}

\def\w{\omega}

\def\s{\sigma}
\def\S{\Sigma}
\def\d{\delta}

\def\k{\kappa}

\def\pd{\partial}
\def\e{\epsilon}

\def\m{\mu}
\def\r{\rho}

\def\L{\Lambda}
\def\cL{{\cal{L}}}
\def\cH{{\cal{H}}}

\def\cB{{\cal{B}}}

\def\cD{{\cal{D}}}

\def\cM{{\cal{M}}}
\def\cZ{{\cal{Z}}}
\def\la{\langle}
\def\ra{\rangle}
\begin{document}
\title{SCHWARZ TYPE TOPOLOGICAL QUANTUM FIELD THEORIES}

\author{R. K. Kaul\thanks{kaul@imsc.res.in},~ T. R. Govindarajan\thanks{trg@imsc.res.in}\\
The Institute of Mathematical Sciences,\\
Chennai 600 113, India,\\ 
\and P. Ramadevi\thanks{ramadevi@phy.iitb.ac.in} \\
Department of Physics, I I T Bombay,\\
Powai, Mumbai 400 076, India}

\maketitle
\centerline{\em An abridged version to appear in the  Encyclopedia of Mathematical Physics}
\centerline{\em to be published by Elsevier}

\section{\bf Introduction}
Topological quantum field theories (TQFT)  provide  powerful tools to probe 
topology of manifolds, specifically in low dimensions\cite{atiyah,
sch, witten1, witten2, kaul}.  
This is achieved by incorporating  very large  gauge symmetries in 
the theory which lead to gauge invariant sectors with only topological 
degrees of freedom.  These theories are of two types: 
(a) {\it Schwarz type} or {\it Chern-Simons type}  and
(b) {\it Witten type} or {\it Cohomological type.} 

In Witten type topological  field theories, action is a BRST exact form,
so is the stress energy tensor $T_{\m\nu}$ so that their functional averages
are zero \cite{witten2}. The topological observables in these theories form 
cohomological classes. In four dimensions, such theories involving 
Yang-Mills gauge fields provide a field theoretic representation 
for Donaldson invariants. 

On the other hand, Schwarz type TQFTs  are described by  local action 
functionals which are 
{\it explicitly}  independent of metric \cite{sch, witten1}. 
The examples of such theories are (i) Chern-Simons (CS) theories and 
(ii) BF theories.

The metric independence of the action $S$ implies that stress energy tensor of
a  TQFT is zero:  $\frac{\d~S}{\d~g_{\m\nu}}~ \equiv ~T_{\m\nu}~=~0$. 
There are no local  propagating degrees of freedom; only degrees of freedom are
topological.  Expectation values of metric independent operators $W$ are
also independent of the metric: $ \frac{\d\la W \ra}{\d~g_{\m\nu}}~=~0$.

Three dimensional Chern-Simons theories are of particular interest, for
these provide a framework for the study of knots and links in any
three-manifold. It was  A.S. Schwarz who first conjectured \cite{sch} that
the well known Jones polynomial may be related to Chern-Simons theory.
In his famous paper, E. Witten\cite{witten1} not only demonstrated this 
connection, but
also set up a general field theoretic framework to study topological
properties of knots and links in any arbitrary three-manifold. In addition 
this framework provides a method of obtaining some new 
manifold invariants. 
Chern-Simons theory is also known to describe 
gravity in three dimensional spacetime \cite{witten3}.

$BF$ theories in three dimensions also provide a field theoretic description
of topological properties of knots and links. 
These theories with bilinear action in fields
in addition can be defined in higher dimensions. 
In particular in $D=4$, BF theory, 
besides describing two-dimensional generalizations of 
knots and links, also provides a field theoretic interpretation  of Donaldson
invariants.  This provides a connection of these theories with
Witten-type TQFTs of Yang-Mills gauge fields. 

Chern-Simons theories in three complex dimensions
described in terms of  holomorphic 1-forms have also been constructed. 
Such a theory on Calabi-Yau spaces can also be interpreted as a string theory 
in terms of a  Witten-type topological field theory
of a sigma model coupled to gravity\cite{witten4}. The observables in
this framework are the
cohomological classes on the moduli space of Riemann surfaces. 
Such holomorphic generalizations of BF theories have also been 
developed \cite{popov}.

In the following we shall survey  three dimensional Chern-Simons theory
as a description of knots/links, indicate how manifold invariants
can be constructed from invariants for framed links, and
also its application to three dimensional gravity.
This will be followed by a brief discussion of BF theories in three 
and four dimensions.

\section{Three-dimensional Chern-Simons theory with \\
gauge group $U(1)$}
The simplest Schwarz type topological field theory is the $U(1)$ Chern-Simons theory  
described by the action:
\be
S_{CS}~=~-\frac{1}{8\pi}\int_{\cM}~A~dA~
\ee
where $A$ is a connection one-form $A~=~A_\m~dx^\m$ and $\cM$ is the three manifold,
which  we shall take to be $S^3$ for the discussion below. 
The action  has no dependence on the metric. Besides being the $U(1)$ gauge
invariant, it is  also general coordinate invariant. 

In quantum CS field theory, we are interested in the functional averages
of the gauge invariant and metric independent functionals $W[A]$:
\be
\la~W[A]~\ra~=~ \frac{1}{\cZ}~\int [\cD A]~W [A]~exp~\{ikS_{CS}\},
~~\cZ~=~\int [\cD A]~exp\{ikS_{CS}\}
\ee

This  theory captures some of the simple, but interesting,
topological properties of knots and links in three dimensions.
For a knot $K$, we associate a  knot operator ~  $\oint_{K} A$
 ~ which is gauge invariant and also does not depend on the metric 
of the three-manifold. Then for  a link made of two knots $ K_1$ 
and $ K_2$, we have the loop
correlation function ~$\la ~\oint_{K_1}A~\oint_{K_2}A~\ra$  which can be
evaluated  in terms of two-point correlator
~ $\la A_\m(x)A_\nu(y)\ra$~  in $R^3$ (with flat metric).
This correlator in Lorentz gauge ($\pd_\m A^\m~=~0$) is:
\beq
\la A_\m(x)A_\nu(y)\ra~=~\frac{i}{k}~\e_{\m\nu\r}~ \frac{(x~-y)^\r}{|x-y|^3}
\eeq
so that for two distinct knots $K_1$ and $K_2$ 
\beq
\la\oint_{K_1}A~\oint_{K_2}~A \ra ~&=&\frac{4\pi i}{k}~\cL(K_1,K_2)
\eeq
where 
\beq
\cL(K_1,K_2) = {\frac{1}{4\pi}}~\oint_{K_1}\oint_{K_2} dx^\m dy^\nu
\e_{\m\nu\r}\frac{(x~-y)^\r}{|x-y|^3}
\eeq
This integral is the well known topological 
invariant called {\it Gauss linking number}\cite{gauss}
of two distinct closed curves. It is an integer  measuring the number 
of times one knot $K_1$ goes through the other knot $K_2$. Linking number 
does not depend on the location, size or shape of the knots. In
electrodynamics, it has 
the physical interpretation  of work done to move a 
monopole around a knot while electric current runs through the other 
knot\cite{maxwell}.

Abelian CS theory also provides a field theoretic representation 
for another topological quantity called 
{\it self-linking number}, also known as {\it framing number}, of the knot. 
It is related to the
functional average of $~\la ~\oint_{K}A~\oint_{K}A~\ra~$ where the two loop
integrals are over the same knot. The coincidence singularity is  avoided
by a topological loop-splitting regularization. 
For a knot  $K$ given  by $x^\m(s)$ parametrized along the length of the 
knot by $s$, we  associate 
another closed  curve $K_f$ given by $y^\m(s)~=~x^\m(s)~+~\e~n^\m(s)$ 
where $\e$ is a small parameter and $n^\m(s)$ is a principal normal 
to the curve at $s$. The coincidence limit is then obtained at the 
end by taking the limit $\e \rightarrow 0$.
Such a limiting procedure is called
{\it framing} and knot $K_f$ is the {\it frame} of knot $K$.
The linking number of the knot $K$ and its frame $K_f$ is  
the self-linking number of the knot:  
\be
S\cL (K, n^\m)~=~ {\frac {1} {4\pi}} {\oint\oint}_{y^\m~=~x^\m~+~\e~n^\m } 
dx^\m~dy^\nu 
~\frac{\e_{\m\nu\r}(x~-~y)^\r}{|x~-~y|^3}
\ee
Hence  coincidence two loop correlator is
\be
\la \oint_K A ~ \oint_K A \ra ~=~ \frac{4\pi i}{k} S\cL(K,n^\m)
\ee
Notice that the self-linking number of a knot  is independent of the 
regularization parameter $\e$, but does
depend on the topological character of the normal vector field $n^\m (s)$.
It is also related to two {\it geometric}
quantities called {\it twist} $T(K)$ and {\it writhe} $ w(K)$ through
a theorem due to Calugareanu:

\be S\cL(K)~=~T(K)~+~\w(K)
\ee
where 
\beq
T(K)~&=&~\frac{1}{2\pi}{\oint_K} ds ~\e_{\m\nu\r}~\frac{dx^\m}{ds}n^\nu 
\frac{dx^\r}{ds}\\
\w (K)~&=&~\frac{1}{4\pi}{\oint_K} ds \oint_K dt ~\e_{\m\nu\r}~\frac{de^\m}{ds}
\frac{de^\nu}{dt}
e^\r
\eeq
Here
\bn e^\m(s,t)~=~\frac{y^\m(t)~-~y^\m(s)}{|y(t)~-~y(s)|}
\en
is a unit map from $K~\otimes~ K \longrightarrow S^2$ and $n^\m(s)$ is a normal 
unit vector field. $T(K)$ and $\w(K)$ are not in general integers and represent the amount of
twist and coiling of the knot. These are not topological invariants
but their sum, self-linking number,  is indeed always an integer and a topological
invariant. 
This result has found interesting application in the studies of
the action of enzymes on circular DNA\cite{crick}.

\section {\bf Non-Abelian Chern-Simons theories}
Non-Abelian CS theories provide far more information about the topological 
properties of the manifolds as well as knots and links in them. 

Non-abelian CS theory in a three manifold $\cM$ (which as in last section we take to be $S^3$)  
is described by the action functional
\be 
S_{CS}~=~\frac{1}{4\pi}\int_{\cM} tr \left(A\wedge dA~+~\frac{2}{3}A\wedge A\wedge A 
\right)
\ee
where $A$ is a gauge field one-form 
which takes its value in the Lie algebra $ \cL \cal G$ of a
compact semi-simple Lie group $ \cal G$. For example, we may take
this group to be $SU(N)$ and $A~=~A^a T^a$ where $T^a$ is
the fundamental $N$ dimensional representation with $ tr~T^a T^b 
~=~ -1/2~\delta^{a b} $. As before the action functional 
is  metric independent and general coordinate invariant. Under 
homotopically non-trivial  gauge transformations this action is not
invariant, but changes by an amount $2\pi n$ where integers $n$ are the winding
numbers characterzing  the gauge transformations which fall in homotopic 
classes given by $\Pi_3 ({\cal G}) = {\cZ}$ for a compact semi-simple group 
$\cal G$. (For a general manifold $\cM$, this change in action is characterised by 
the homotopic class of maps $\cM \longrightarrow {\cal G}$.) 
However, for quantum theory what is relevant is $exp[ikS_{CS}]$
which is invariant even under homotopically non-trivial gauge transformations
provided the coupling $k$ takes integer values \cite{deser1}. So for integer
$k$, the quantum field theory we discuss here is gauge invariant.

The topological operators are the Wilson loop operators for an 
oriented  knot $K$:
\be
W_R[K]~=~tr~P~exp~\oint_K A_R
\ee
where $A_R = A^a T^a_R$ with $T^a_R$ as the representation matrices of 
a finite dimensional representation $R$
of the $\cL \cal G$. $P$ stands for the path ordering of the exponential. 
The observable Wilson link operator  for a link $L~=~\bigcup_1^n K_i$,
carrying representations $R_i$ on the respective component knots, is
\be W_{R_1R_2\cdots R_n}\left[L\right]~=~ \prod_1^n ~W_{R_i}[K_i]
\ee
Expectation values of these operators are the functional averages:
\be 
V_{R_1,R_2\cdots R_n}[L]~=~\frac{\int [\cD A]~ W_{R_1R_2\cdots R_n}[L]~
exp[ikS_{CS}]} {\int [\cD A]~exp[ikS_{CS}]}
\ee
The measure $[\cD A]$ has to be metric independent. 
These expectation values depend only on the isotopy of the link $L$ and also
on the set of  the representations $\{ R_i \}$. These
can be evaluated in principle nonperturbatively. For example when 
$\cL{\cal G} ~=~ su(N)$ and  each the component knot of the links carries  
the fundamental $N$ dimensional representation, the Wilson link 
expectation values satisfy a recursion relation involving three link diagrams 
which are identical except for one crossing where they differ as over
crossing $(L_{+})$, under crossing $(L_{-})$ and no crossing $(L_0)$ as shown 
in the figure below. 
\vskip.4cm
\centerline{\includegraphics[scale=.5]{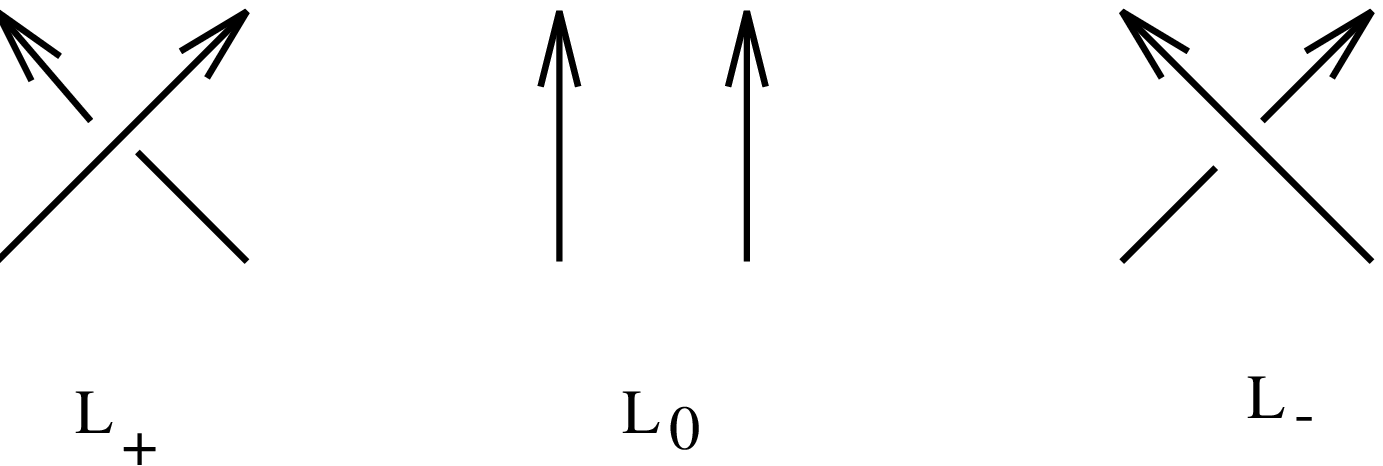}}
\vskip.1cm
\noindent The expectation values of these links are related 
as \cite{witten1}:
\be
q^{\frac{N}{2}}~V_N[L_+]~-~q^{-\frac{N}{2}}~V_N[L_-]~=~\left(q^{\frac{1}{2}}~-~q^{-\frac{1}{2}}\right)
V_N[L_0]
\ee
where $q~=~exp\left(\frac{2\pi i}{k~+~N}\right)$. This is precisely the 
well known skein relation for the HOMFLY polynomial. The famous Jones
one-variable polynomial (whose two-variable generalization is the
HOMFLY polynomial), correspond to the case of spin-$1/2$ representation 
of $SU(2)$ CS theory:  
\be
V_2[L]~=~ Jones~Polynomial~[L]
\ee
upto an overall normalization.
These skein relations are sufficient to find recursively all the 
expectation values of 
links with only fundamental representation on the components. To obtain 
link invariants for any other representation more general methods have 
to be developed. A complete and explicit solution of the Chern-Simons
field theory is thus obtained.
One such method has been presented in ref.\cite{kaulcmp}.
The method makes use of the following important theorem:
 
\vspace{0.1cm} 

\noindent{\bf Theorem:} Chern-Simons theory on a three-manifold $\cM$ 
with boundary $\S$
is described by a WZNW (Wess-Zumino-Novikov-Witten) conformal field theory
on the boundary. 
\vskip.1cm
\centerline{\includegraphics[scale=.4]{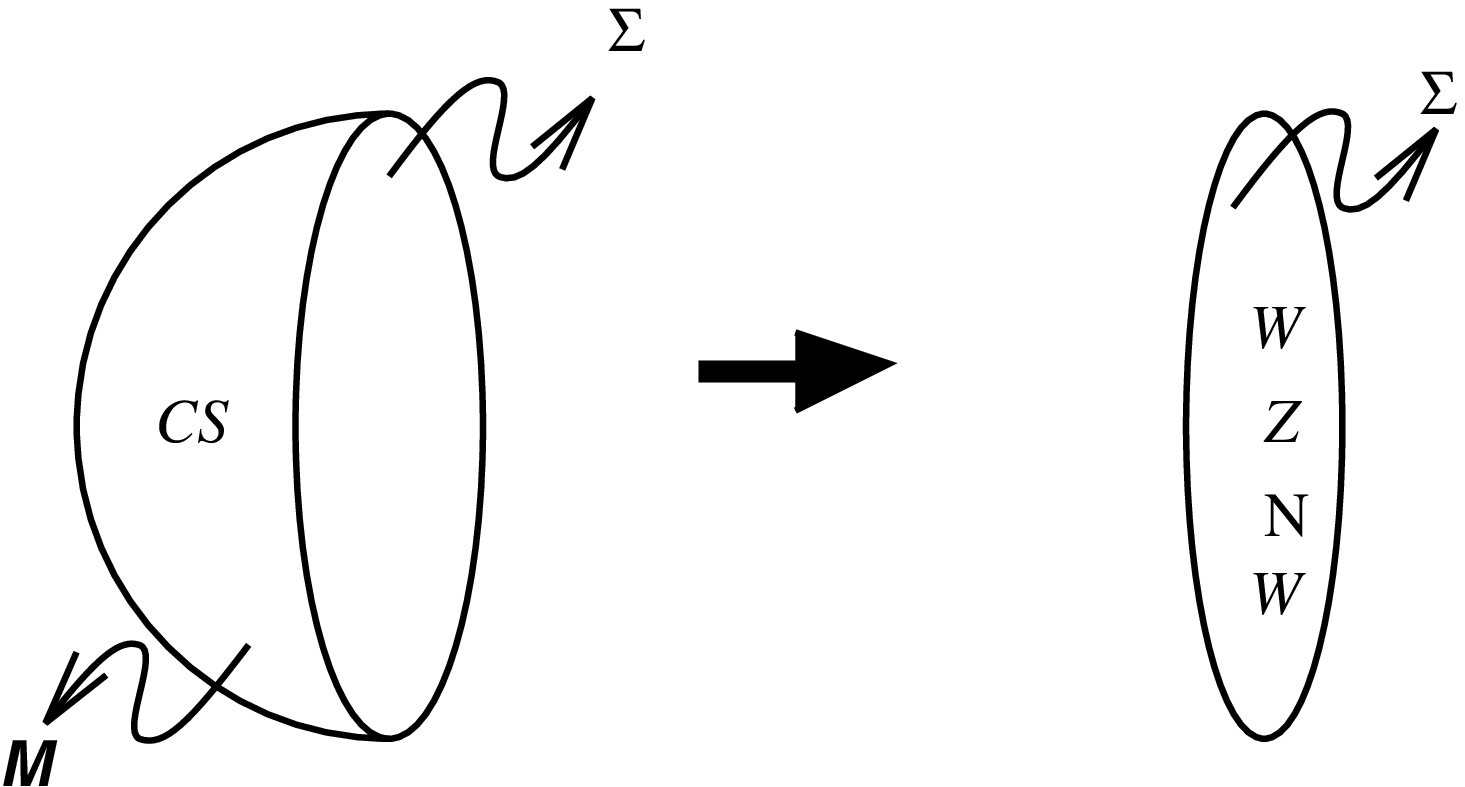}}
\vskip.2cm
Using the same identification  the functional  average  
for Wilson lines 
ending at $n$ points on the boundary $\S$ is obtained from WZNW 
field theory  on the boundary with $n$ punctures
carrying representations $R_i$:
\vskip.2cm
\centerline{\includegraphics[scale=.35]{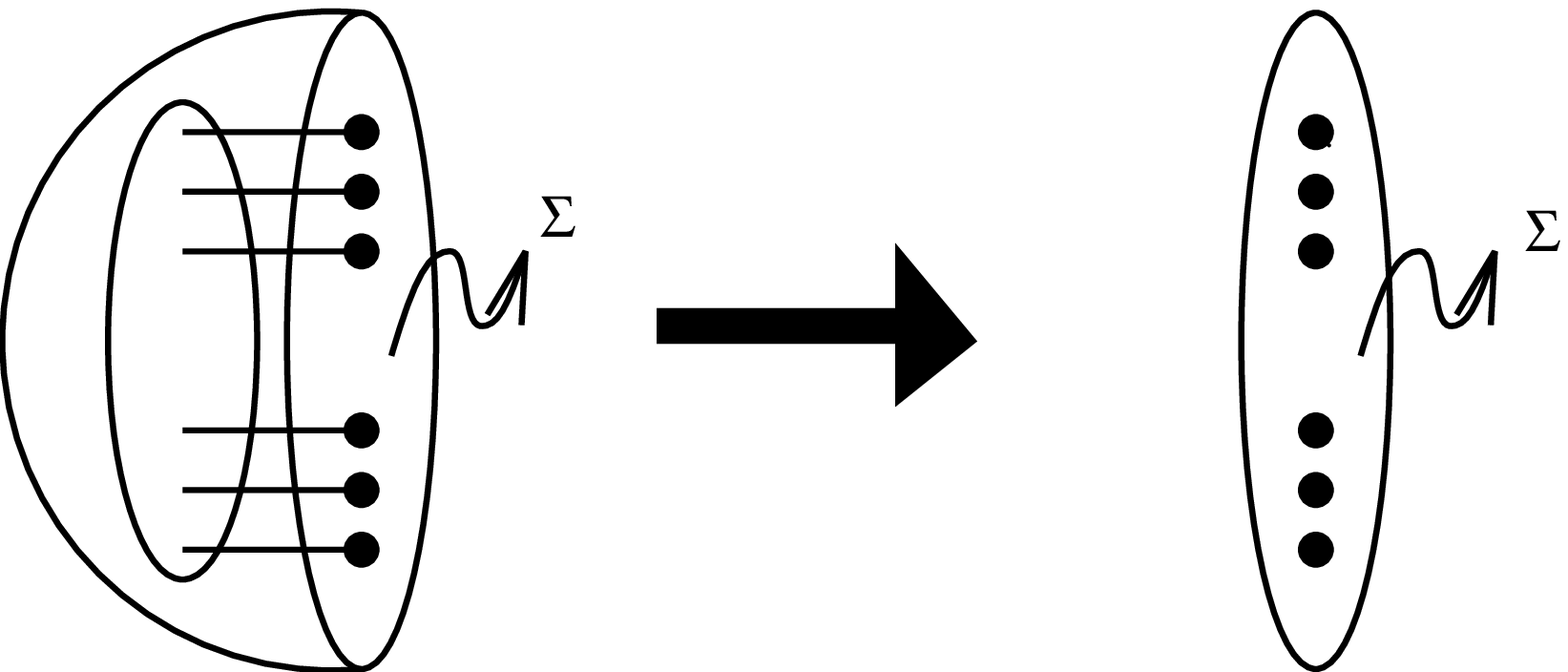}}
\vskip.2cm
We can represent CS functional integral as a vector in the 
Hilbert space $\cH$ associated with 
the $n$-point vacuum expectation values of the primary fields in
WZNW conformal field theory on the boundary $\S$\cite{witten1}.
Next, to obtain a complete and explicit nonperturbative solution of the 
CS theory, the theory of knots and links  and their connection to  
braids is invoked \cite{kaulcmp,ram}. 

\subsection{\bf Knots/links and their connection to braids} 

Braids have an intimate connection with knots and links. This connection
is summarized  as follows:

\begin{enumerate}
\item An $n-$braid is a collection of non intersecting strands connecting $n$ points 
on a horizontal rod
to $n$ points on another horizontal rod below strictly excluding any backward 
traversing of the strands. A general braid can be written as a word in terms of elementary
braid generators. 
\item We associate representations $R_i$ of the group with the strands as their 
colors. We also put an orientation on each strand.
When all the representations are identical and also all strands are
oriented in the same direction, we get ordinary braids, 
otherwise we get colored oriented braids.
\item The colored oriented braids form a groupoid  where product 
of the different braids is obtained by joining them with
both colors and orientations matching on the joined strands.
Unoriented monochromatic braids form a group.
\item A  knot/link can be formed from a given braid by a process called
platting. We connect adjacent strands namely the $(2i+1)^{th}$ strand to 
$2i^{th}$ strand carrying the same color and opposite orientations in both 
the rods of an even-strand braid as shown in figure (a) below. 
\vskip.3cm
\centerline{\includegraphics[scale=.4]{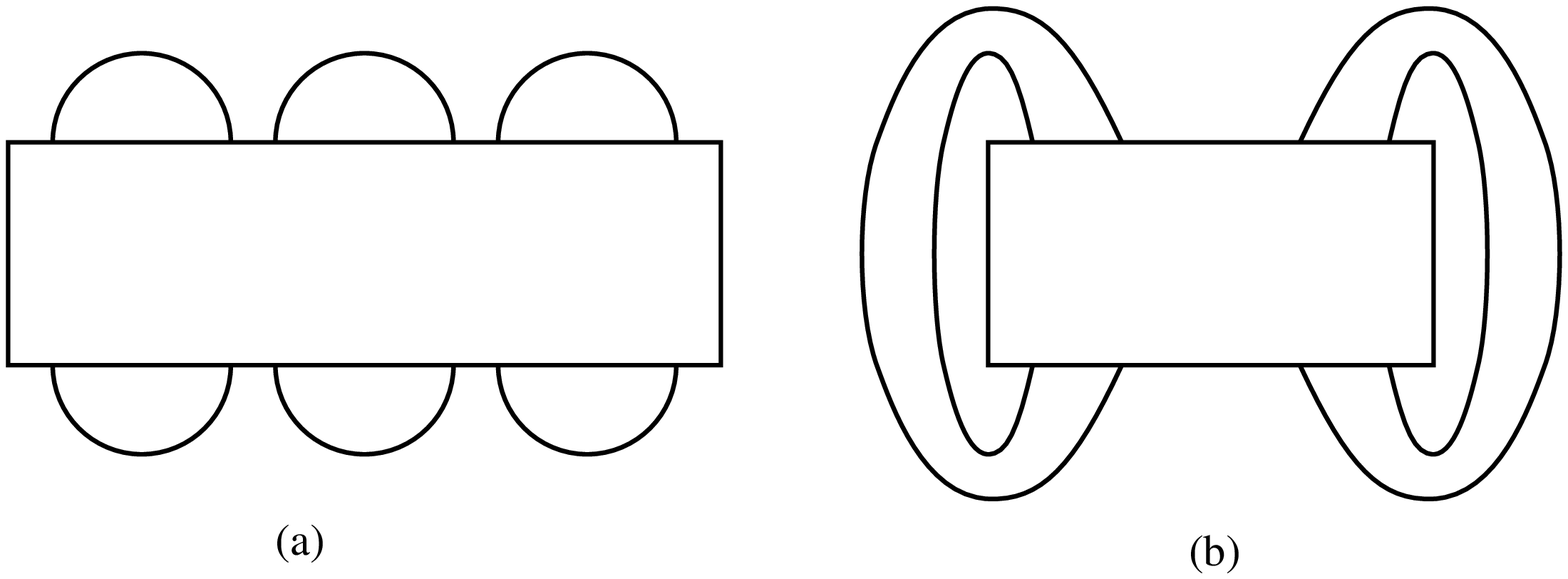}}
\vskip.2cm
There is a theorem due to Birman which states that all colored 
oriented knots/links
can be obtained through platting \cite{birman}. This construction 
is not unique.
\item There is another construction associated with braids which relates
them
to knots and links. We obtain a closure of a braid by connecting the ends
of the first, second, third, .....strands from above to those of the
respective first, second, third, ... strands from below as shown in the
figure (b) above. There is theorem due to Alexander\cite{birman} which states that
any knot or link can be obtained as a closure of a braid, though 
again not uniquely. 

\end{enumerate}

\subsection {Link invariants}

This connection of braids to knots and links can be used to construct
link invariants, say in $S^3$,  from the Chern-Simons theories.
To do so,  from the three-manifold $S^3$ two non intersecting 3-balls are removed 
to obtain a manifold with  two $S^2$ boundaries. Then we can arrange $2n$ 
Wilson line of, say $SU(N)$ Chern-Simons theory, as  a $2n$-strand oriented
braid  carrying representations $R_i$ in this manifold. The
CS functional integral over this manifold is a state in the tensor 
product of the Hilbert spaces  $\cH_1\otimes \cH_2$
associated with the conformal field theory on the two  boundaries.
The two boundaries have  $2n$ punctures   carrying the set 
of representations $\{R_i\}$ and $\{R'_i\}$ respectively, the two sets being
permutations of each other.
This state can be expanded in terms of some convenient
basis given by the conformal blocks for the $2n$-point correlation 
functions of the $SU(N)_k$ WZNW conformal field theory. 
The duality of these correlation functions represents the transformation between different 
bases for the Hilbert space. Their monodromy properties allow us to write
down representations of the braid generators. Since an arbitrary braid
is just a word in terms of these generators, this 
construction provides us a matrix representation $\cB(\{R_i\},\{R'_j\})$ 
for the colored-oriented braid in the manifold with two $S^2$ boundaries.
Then we plat this braid by gluing two balls $B_1$ and $B_2$ with Wilson lines as shown in the figure:
\vskip.3cm
\centerline{\includegraphics[scale=.4]{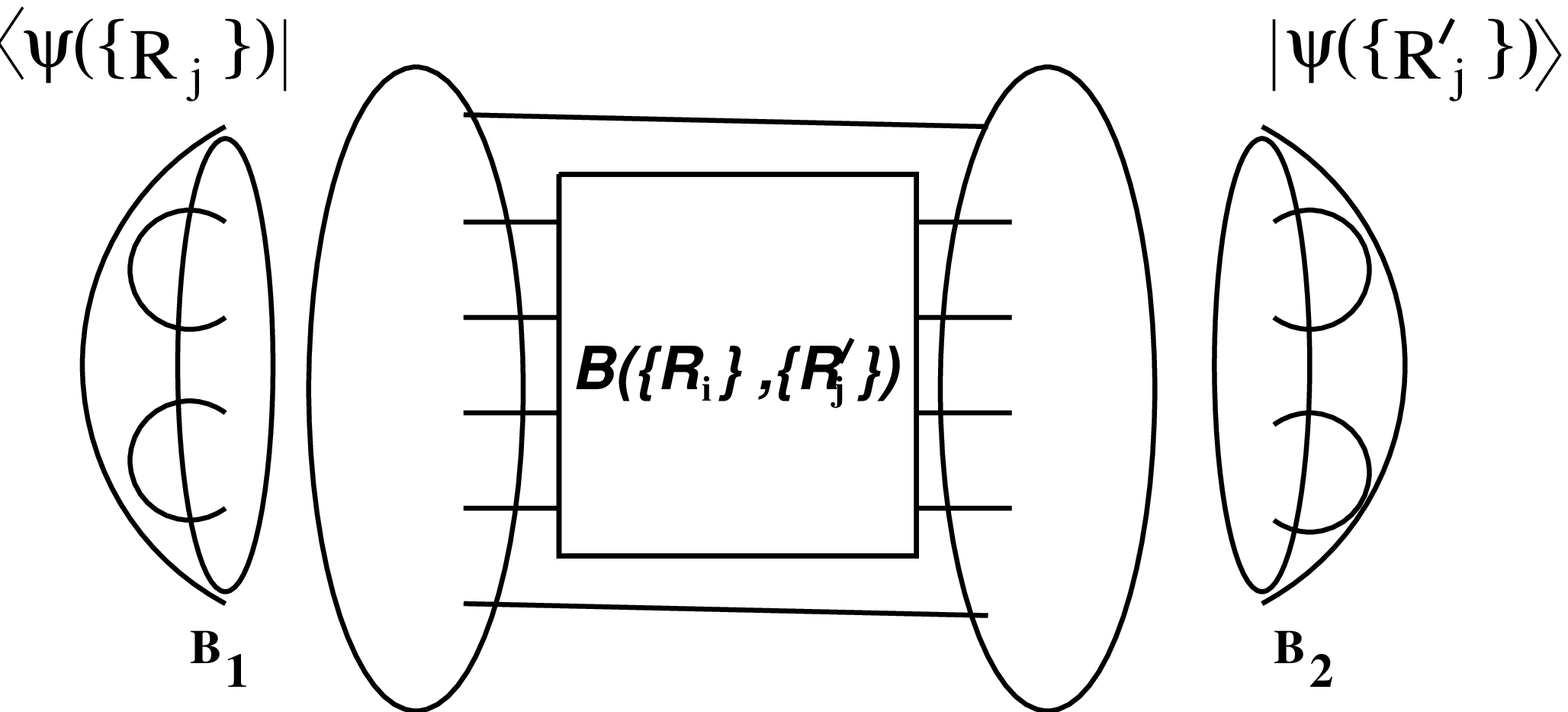}}
\vskip.1cm
Each of the two  caps  again represents 
a state $|\psi(\{R_j\})\ra$ in the Hilbert space associated with the conformal 
field theory on the  punctured boundary ($S^2$).  Platting of the braid then
simply is the matrix element of the braid representation
$\cB(\{R_i\},\{R'_j\})$ with respect to these states $|\psi(\{R_i\})\ra$
and $|\psi(\{R'_j\})\ra$ corresponding to the
two caps $B_1, B_2$. Thus for a link in $S^3$ the invariant is given by the following
proposition: 

\vspace{0.2cm}

\noindent{\bf Theorem:} The vacuum expectation value of Wilson loop operator 
of a link $L$ constructed from platting of a  
colored oriented $2n-$braid with representation $\cB(\{R_i\},\{R'_j\})$ 
is given by:
\be
V[L]~=~\la~\psi(\{R_i\})|\cB(\{R_i\},\{R'_j\})|\psi(\{R'_j\})\ra
\ee  

For detailed proof of this theorem for gauge group $SU(2)$ see 
ref. \cite{kaulcmp}.
This theorem can be used to calculate the invariant for any arbitrary link.
For an unknot $U$ carrying a fundamental N dimensional representation in an
$SU(N)$ CS theory, the knot invariant is:
\be
V_N[U]~=~\left[N\right], \qquad  where ~~[N]~=~\frac{q^{\frac{N}{2}}~-~q^{-\frac{N}{2}}}
{q^{\frac{1}{2}}~-~q^{-\frac{1}{2}}}
\ee

The Wilson link expectation values calculated this way depend on the 
regularization, i.e., the definition of framing used in 
defining the coincident loop correlators. One such regularization usually
used is the  {\it standard} framing where the frame for every knot
is so chosen that its self-linking number is zero.

The procedure outlined here has been used for  explicit computations
of knot/link invariants\cite{kaulcmp,ram}.  This has led to answers to several questions 
of knot theory\cite{ram1}. One such question relates to distinguishing chirality
of knots.  For example,  among the knots 
in knot tables\cite{rolfsen},  the knots 
$9_{42}$ and $10_{71}$ and their respective mirror images are not distinguished
by any older polynomial invariants including Jones and HOMFLY  polynomials.
But  expectation value of the Wilson loop operator with spin $3/2$ 
in $SU(2)$ CS theory does distinguish them. Though knot invariants
obtained from CS theories are sensitive to the chirality of many
knots, yet they do not distinguish chirality of {\it all} knots. 
Inspite of the success of CS theory
in its application to knot theory it fails to distinguish 
a class of links known as {\it mutant} links \cite{ram1}. A  mutant link
is obtained by removing a portion of weaving pattern in a link which is
rotated about any one of three orthogonal axes by an amount $\pi$ 
and then glued back in to the link.

The CS invariants of knots and links  can also be used 
to construct special three-manifold invariants. 
Hence CS theory provides an important tool to study these.

\section{Three-manifold invariants from Chern-Simons theory}
Classification of three dimensional manifolds is one of the 
challenging problems in mathematics. Interestingly, different
three-manifolds can be constructed through a procedure called
{\it surgery} of {\it framed knots and links} in three-sphere $S^3$
(Lickorish-Wallace theorem)\cite {lic, kaul}. 
However, the surgery construction relating
the framed knots and links to the corresponding three-manifold is not unique.
That is, there are many framed knots and links which
give the same manifold. 
However rules of this equivalence  are known: these are called Kirby 
moves\cite {kir}. 

Classification of  three-manifolds would involve finding
a method of associating a {\it quantity}  with the manifold 
obtained by surgery on the corresponding framed knot/link on $S^3$. If the
Kirby moves on the framed knot/link leave this quantity  
unchanged, then it is a
{\it three-manifold invariant}. Knot/Link invariants of non-abelian
CS theories provide a method of finding such three-manifold 
invariants \cite {kaul, kaul1}. Equivalently, this procedure
gives an algebraic meaning to the surgery construction of three-manifolds.

\subsection {Surgery of framed knots/links and Kirby moves}

\noindent
 As discussed earlier, {\it frame} of a knot $K$ is an associated closed curve 
$K_f$ going along the 
length of the knot wrapping around it certain number of times.
Self-linking number (also called framing number)
is equal to the linking number of the knot with its frame.
There are several ways of fixing this framing. The  {\it standard} 
framing is one in which the frame number of the knot, that is, the linking
number of the knot and its frame is zero. On the other hand, {\it
vertical} framing is obtained by choosing the frame vertically above 
the knot projected on to a plane. In such a frame, the framing number 
of a  knot is the same as  its {\it crossing number}.
In constructing the three-manifold invariants from Chern-Simons theories,
we need vertical framing. The framing number may be denoted by writing the 
integer by the side of knot. We denote   a framed $r$-component
link by $[L, {\bf f}]$ where framing 
${\bf f}=\left (n(1), n(2) \ldots n(r)\right)$ is a set of integers 
denoting the framing number of component knots $K_1, K_2 \ldots K_r$ 
in the link $L$. 

According to the Lickorish-Wallace theorem,  surgery over links 
with vertical framing in $S^3$  yields all the three-manifolds.
This surgery is performed in the following way.

Take a framed $r-$component link $[L, {\bf f}]$ in $S^3$. Thicken 
the component knots $K_1, K_2 \ldots K_r$
such that the  solid tubes $N_1, N_2, \ldots N_r$ so obtained are 
non-intersecting. Then the compliment $S^3 - \left(N_1+N_2 
\ldots + N_r \right)$ will have $r$ toral boundaries. On the $i^{th}$
 toral boundary,  we imagine an  appropriate curve winding $n(i)$ times
around the meridian and once along the longitude.
Perform a modular transformation so that this curve bounds a disc.
This construction is done with each of the toral boundaries. The tubes
$N_1, N_2, \ldots N_r$  are 
then glued back in to the respective gaps. This surgery thus
yields a new three-manifold.
This construction is not unique. 
The rules of equivalence for surgery on framed knots/links in $S^3$ are 
the two Kirby moves\cite{kir}. 

\vspace{0.2cm}

{\bf Kirby move I}: ~ Take an arbitrary $r$-component
framed link $[L, {\bf f}]$ in $S^3$ and consider a curve $C$ with framing number 
$+1$ going around the unlinked strands of $L$ as shown in Figure 6(a).
We refer to this $(r+1)$-component link as $H[X]$, where $X$ represents
a weaving pattern of the strands.
Kirby move I constitutes of twisting the disc enclosed by $C$ in 
the {\it{clock-wise}} direction
from below by an amount $2\pi$. This twisting thereby introduces
new crossings between the curve $C$ and  the strands enclosed by it.
Then the curve $C$ is removed giving us a new $r$-component link  $U[X]$
as shown in Figure 6(b). The framing numbers $n'(i)$ 
of the component knots in link $U[X]$ are related to the framing
number $n(i)$ of the framed link $[L, {\bf f}]$  
as $n'(i)= n(i) - \left(\cL (K_i, C)\right)^2$, where $\cL (K_i, C)$
is the linking number of  knot $K_i$ and  closed curve $C$.
The surgery of the two framed links in the figure below  
will give the same three manifold.
\vskip.3cm
\centerline{\includegraphics[scale=.3]{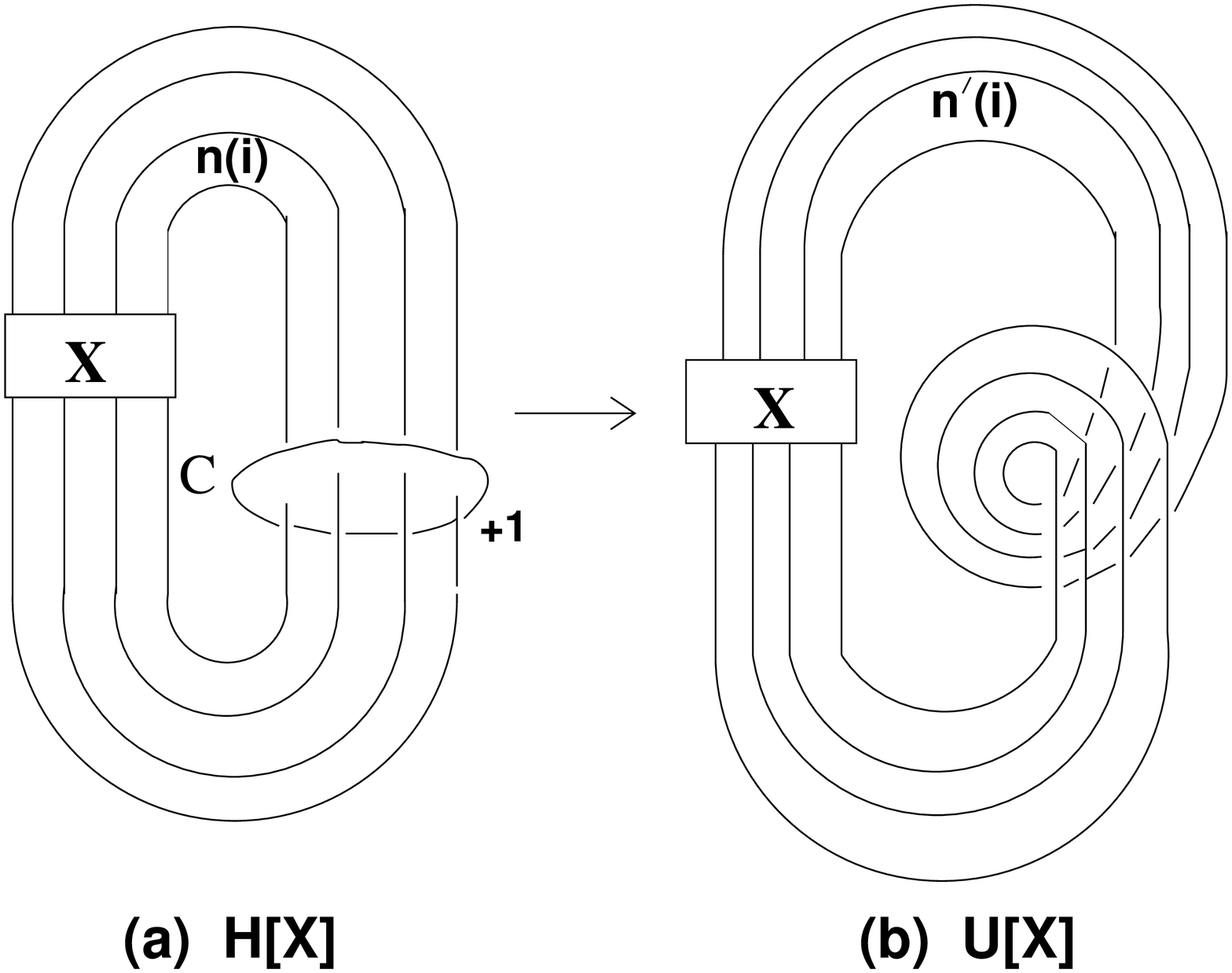}}  
\vskip.2cm
Inverse Kirby move I involves removal of a curve $C$ with framing number $-1$
(instead of $+1$) after making one complete {\it{anti-clockwise}} twist 
from below 
on the disc enclosed by $C$. In the process the unlinked strands get twisted 
in the anti clockwise direction leading to changed framing numbers $n'(i)= 
n(i) + \left(\cL (K_i,C) \right)^2$ 
of the component knots $K_i$ 

\vspace{.2cm}
 
{\bf Kirby move II}: ~ This move consists of removing
a disjoint unknot $C$ with framing $-1$ from  framed link $[L,{\bf f}]$
without changing the rest of the link as shown in the figure below.  
The surgery of two framed links in this figure 
give the same three-manifold.
\vskip.2cm
\centerline{{\includegraphics[scale=.5]{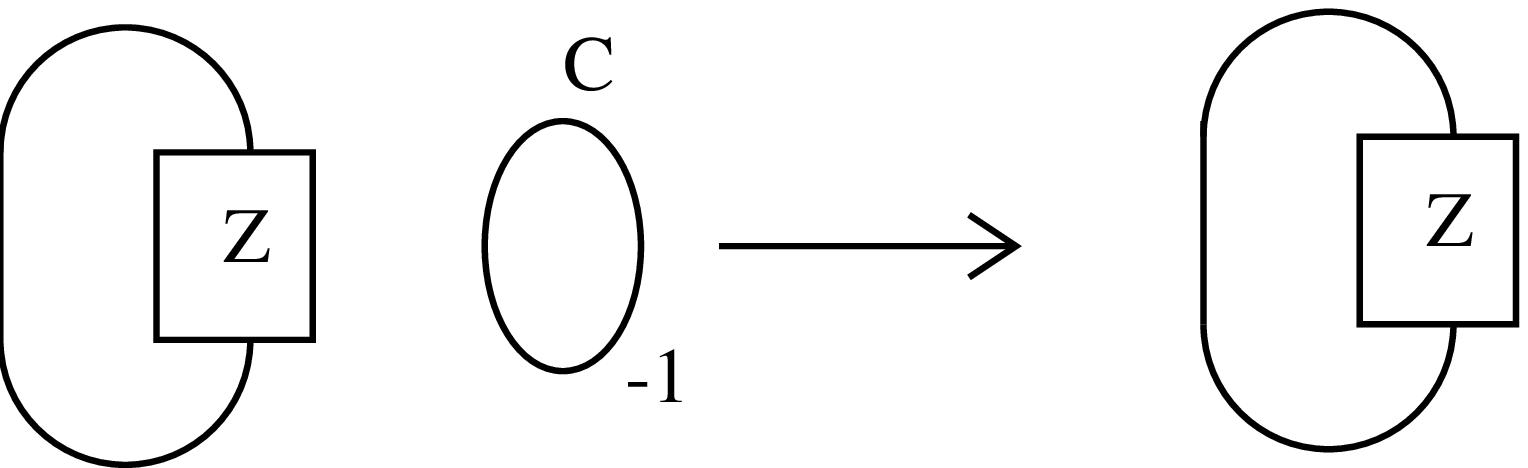}}}
\vskip.2cm
Inverse Kirby move II involves removal of a disjoint unknot with framing
$+1$ (instead of $-1$) from a framed link.

\subsection{Three-manifold invariants}

Now a  three-manifold invariant can be
constructed by an appropriate combination of the invariants of 
framed links in such a way that this algebraic expression is unchanged under
the Kirby moves I and II. What we need  for this purpose  are the invariants
for links with vertical framing in $S^3$.

Let $\cM$ be the three manifold obtained from surgery of a 
$r$-component framed link $[L,{\bf f}]$ in $S^3$.
Then a three-manifold invariant $\hat F^{(\cal G)}[\cM]$ 
for $\cM$ is given as a linear combination  of the framed link 
invariants $V^{(\cal G)}_{R_1, R_2, \ldots
R_r}[L,{\bf f}]$,  with representations $R_1,R_2,\ldots R_r$ living
on component knots,  obtained from Chern-Simons theory based on any compact 
semi-simple group ${\cal G}$ \cite{kaul, kaul1}:
\begin{equation}
\hat F^{(\cal G)}[\cM] = \alpha^{-\sigma[L, {\bf f}]} \sum_{R_1,R_2, \ldots R_r} 
~\left(\prod_{i=1}^r
\mu_{R_i}\right)~V^{(\cal G)}_{R_1, R_2, \ldots R_r}[L, {\bf f}]~,
\label{invariant}
\end{equation}
Here $\sigma[L, {\bf f}]$ is the signature of the linking
matrix and 
\begin{equation}
\mu_{R_i} = S_{0 R_i}~,~ \alpha= e^{i \pi c / 4}~,
\end{equation}
where $c$ is the central charge of the associated WZNW conformal field theory
and $S_{0R_i}$ denotes the matrix element of the modular matrix $S$.
General $S$-matrix elements for any compact group are given by\cite{smatrix}
\begin{eqnarray}
S_{R_1 R_2} &=& (-i)^{d-r \over 2}
\vert {L_{\omega} / L} \vert^{-{1 \over 2}}
\left(k + C_v \right)^{-{1 \over 2}}
\sum_{\omega \in W} \epsilon (\omega)\nonumber\\
~&~& \exp \left( { -2 \pi i \over k+C_v}
(\omega (\Lambda_{R_1} + \rho), \Lambda_{R_2} + \rho) \right)
\label{smatrix}
\end{eqnarray}
where $W$ denotes the Weyl group and its elements $\omega$
are words constructed using the generator $s_{\alpha_i}$ --
that is, $\omega = \prod_i s_{\alpha_i}$ and
$\epsilon(\omega) = (-1)^{\ell (\omega)}$ with
$\ell (\omega)$ as  length of the word. Here $\Lambda_{R_i}$'s
denotes the highest weights of the representations $R_i$'s
and $\rho$ is the Weyl vector.
The action of the Weyl generator $s_{\alpha}$ on a weight $\Lambda_R$
is: ~
$
s_{\alpha} (\Lambda_R) = \Lambda_R - 2 \alpha {(\Lambda_R, \alpha)
\over (\alpha, \alpha)}~,
$
and ~$\vert {L_{\omega} /  L }\vert$ is the ratio of
weight and  co-root lattices
(equal to the determinant of
the cartan matrix for simply laced algebras).

It is important to stress that the expression $\hat F^{(\cal G)}[\cM]$ 
can be shown to be unchanged under both Kirby moves I and II (for detailed 
proof, see refs. \cite{kaul, kaul1}).  Notice that
{\it for every compact gauge group, we have a new three-manifold invariant}.

\vspace{0.2cm}

\noindent
{\bf Few examples of three-manifolds}

\vspace{0.1cm}

\noindent We now list  the algebraic expressions of this invariant
calculated explicitly from formula in Eq.[\ref{invariant}] for a 
few three-manifolds in the table below. 
We have indicated  the framed links in $S^3$ which on  
surgery yield the corresponding three-manifolds $\cM$. ${\cal L}[p,q]$
stands for Lens spaces of the  type $(p,q)$ and 
$C_R$ is the quadratic Casimir invariant for representation $R$ of the Lie
algebra of the gauge group $\cal G$.

\vspace{0.5cm}

\begin{tabular}{||l|l|l||}  \hline
~~Framed link& $~~~\cM$ & $~~\hat F^{(\cal G)}[\cM]$ \\\hline 
&&\\
~~Unknot with zero framing ~~~&$~S^2 \times S^1~~$&$~ 1/S_{00}$\\
&&\\
~~Unknot with framing $\pm 1$ &$~S^3$ &$ ~1$\\
&&\\
~~Unknot with framing $+2$&$~RP^3$&$~\alpha^{-1}{\sum_R}~{S_{0R}q^{2C_R}
S_{0R} \over S_{00}}$\\
&&\\
~~Unknot with framing $+p$&$~{\cal L}[p, 1]$& 
$~\alpha^{-1}\sum_R~{S_{0R}q^{pC_R}S_{0R} \over S_{00}}$~\\ 
&&\\ \hline
\end{tabular}

\vspace{0.5cm}

 For a Chern-Simons theory in a manifold $\cM$, the partition function
is also an invariant characterizing the three-manifold $\cM$.
This has been calculated for several manifolds by different methods\cite{jeff}.
From the expressions for the invariant $\hat F^{(\cal G)}[\cM]$ 
for various manifolds listed above, it appears that this  invariant 
is related to the Chern-Simons partition function $Z^{(\cal G)}[\cM]$ as follows:
\begin{equation}
\hat F^{(\cal G)} [\cM] = {Z^{(\cal G)}[\cM] \over S_{00}}
\end{equation}
For the gauge group $SU(2)$, this relationship has been established
in ref. \cite{ram2}. So the method of constructing three-manifold
invariants above can also be used to calculate the partition function of
Chern-Simons theories.

\section{\bf 3D gravity and CS theory}
Three dimensional Chern-Simons theory also provides a description
of gravity. The  3D gravity  action including the cosmological constant 
$\L~=~\pm 1/ \ell^2$ is:
\be
S~=~\frac{1}{16\pi G}\int_\cM ~d^3x~\sqrt{-g}\left( R~-~2\L\right)
\ee
G is the Newton's constant, $g_{\m\nu}$ is the 
metric on the three manifold $\cM$ and R is scalar curvature. These
theories with cosmological constant were first discussed in 
ref.\cite{deser2}. The Einstein's field equations are:
\be
R_{\m\nu}~-~\frac{1}{2}g_{\m\nu}~R~+~\L~g_{\m\nu}~=~0
\ee
In three dimensions the solutions of these equations have a constant 
positive (negative)
curvature if $\L$ is positive (negative). It is also well known that 
there are no dynamical degrees of freedom for gravity in dimensions 
$D\leq 3$; it is  indeed described by topological field theories.
The gravity action above can be rewritten as 
a Chern-Simons gauge theory \cite{witten3}  in first
order formulation. For triads $e_\m^a$  and spin connection $\w_\m^a$ 
of Euclidean gravity, we define one-forms 
~$e=e_\m^a~T^a~dx^\m,$ ~  $\w=\w_\m^a~T^a~dx^\m$, 
which have values in the Lie-algebra of $SU(2)$ whose generators are 
~$T^a=i\s^a/2$~ with ~$\s^a$~ as three Pauli matrices. In terms of these  
we define two gauge field one-forms $A$ and $\bar{A}$ as:
\bn
A~=~\left(\frac{i~e}{\ell}~+~\w\right),  ~~~~~~~~ 
\bar{A}~=~\left(\frac{i~e}{\ell}~-~\w\right) 
\en
Then the Euclidean gravity action in terms of $A$ and $\bar{A}$ is:
\be
S~=~k S_{CS}\left[A\right]~-~k S_{CS}\left[\bar{A}\right]\label{grav1}
\ee
where the CS coupling constant $k~=~\ell/(4G)$  for negative
cosmological constant $\L = - 1/ \ell^2$. The gauge group
for this theory is $SL(2,C)$.
Infinitesimal diffeomorphisms are described by  field dependent gauge transformations. 
The corresponding gauge group  for Minkowski gravity with negative cosmological 
constant $\L$ is $SO(2,R)\otimes SO(2,R)$. For positive $\L$ one gets $SO(3,1)$ and $SO(4)$ for 
Minkowski and Eucildean metrics respectively.
For $\L~=~0$ we have $ISO(2,1)~ (ISO(3))$ as the gauge group
for Minkowski (Euclidean) gravity. Hence the sign of cosmological
constant determines the gauge group of the CS theory. 

The identification of 3D gravity with CS theory can be used with some
advantage to find the partition function for a black hole in 3D 
gravity with negative cosmological constant. This in turn yields an expression
for entropy of the black hole. 

\subsection{\bf BTZ black hole and its partition function}
Only for  negative $\L$ we have a black hole solution of the Einstein's
equations. This solution, known as the BTZ black hole\cite{btz}, 
in Euclidean gravity  is given by the metric:
\bn
ds_E^2~&=&\left(-M~+~\frac{r^2}{l^2} ~-~\frac{J^2}{4r^2}\right)d\tau^2~\cr
&+&
\left(-M~+~\frac{r^2}{l^2}~-~\frac{J^2}{4r^2}\right)^{-1}dr^2~+~r^2(d\phi~-~\frac{J}{2r}d\tau)^2
\en
It is specified by two parameters $M$ and $J$ (the mass and 
the angular momentum). By a 
coordinate transformation this metric can be rewritten as:
\bn
ds_E^2~=~\frac{l^2}{z^2}(dx^2~+~dy^2~+~dz^2), \qquad z~> 0
\en
This is the 3D upper half hyperbolic space and can be rewritten using spherical polar 
coordinates as:
\be
ds_E^2~=~\frac{l^2}{R^2\sin^2\chi}(dR^2~+~R^2d\theta^2~+~R^2\sin^2\theta d\chi^2)
\ee
We have the identifications 
$\left(R,~\theta,~\chi\right) \sim\left(R~exp\{2\pi r_+/l\},
~\theta +\{2\pi r_-/l\}, ~\chi \right)$ where $r_+ $ and $r_-$ are the 
outer and inner horizon radii respectively.
It is clear from this identification that the metric 
topologically corresponds to a 
solid torus with a boundary. The functional integral over this manifold
represents a state in the Hilbert space
specified by the mass and the angular momentum.  It is the micro canonical 
ensemble partition function and its logarithm is  
entropy of  the back hole. 

To evaluate this partition function, the connection one-form is kept 
at a constant value on the toroidal boundary through a  gauge
transformation. We define local coordinates on the torus boundary $z~=~x~+~\tau y$ such that
~$\int_a~dz~=~1, ~  \int_b~dz~=~\tau$, 
~where $a~(b)$  stands for the contractible (non-contractible) cycle of the solid torus
and $\tau= \tau_1 + i\tau_2$ is the modular parameter of the boundary torus. Then the connection 
describing the black hole is:
\be
A~=~\left(\frac{-i\pi~\tilde{u}}{\tau_2}~d\bar{z}~+~\frac{i\pi~u}{\tau_2}~dz\right)~T_3
\ee
where $u$ and $\tilde{u}$ are canonically conjugate  with 
the commutation relation given by:
\be
\left[\tilde{u},~u\right]~=~\frac{2\tau_2}{\pi(k+2)}
\ee
These are related to black hole parameters through the holonomies of 
the gauge field $A$ around the contractible
and non-contractible cycles: 
\beq
u~=&~-\frac{i}{2\pi}\left(-i\Theta \tau~+~\frac{2\pi(r_+~ + ~i|r_-|)}{l}\right),\cr
\tilde{u}~=&~-\frac{i}{2\pi}\left(-i\Theta \bar{\tau}~+~\frac{2\pi(r_+~+~i|r_-|)}{l}\right)
\eeq
For a classical black hole solution $\Theta~=~2\pi$.

For a fixed value of connection, namely $u$, the functional integral is 
described by a state $\psi_0$ with no Wilson line in the bulk. 
The states with Wilson line carrying spin $j/2$
are given by \cite{lab}: 

\be
\psi_j(u,\tau)~=~exp~\left\{ {\frac{\pi k}{4\tau_2}}~u^2\right\}\chi_j(u,\tau)
\ee
where $\chi_j$ are the Weyl-Kac characters for affine $su(2)$:

\bn
\chi_{j}(u, \tau)~ =~ \frac{\Theta_{j +1}^{(k+2)}(u, \tau, 0)~-~\Theta_{-j -1}^{(k+2)}(u, \tau, 0)}
{\Theta_{1}^{2}(u, \tau, 0)~-~\Theta_{-1}^{2}(u, \tau, 0)}
\en

\noindent where $ \Theta $ functions are defined by:

\bn
\Theta_{\mu}^{k}(u,\tau,0)~=~\sum_{n \in \cal Z} \exp \Big\{2 \pi i k 
\left[(n + \frac{\mu} {2 k})^2 \tau ~+~(n + \frac{\mu}{2 k}) u \right]\Big\}
\en
Given the collection of states $\psi_j$ we write the partition function by choosing 
an appropriate ensemble by fixing the mass and angular momentum.
This black hole partition function is:
\be
Z_{BH} =~\int d\mu(\tau, \bar \tau)\left|~\sum_{j=0}^{k}
~(\psi_{j}(0, \tau))^{*} ~\psi_{j}(u, \tau)~\right|^2
\ee
where modular invariant measure  is
~$d\m(\tau,\bar{\tau})~=~{d\tau~d\bar{\tau}}/{\tau_2^2}$. ~ 
This integral can be worked out for large black hole mass 
and zero angular momentum in  saddle point approximation. 
The computation is  done in ref.\cite{ours} and yields:
\be
Z_{BH}~=~\frac{l^2}{r_+^2}~\sqrt{\frac{8r_+G}{\pi l^2}}~\exp{\left(\frac{2\pi r_+}{4G}\right)}~+\cdots
\ee
This gives not only the leading  Bekenstein-Hawking behavior of 
the black hole entropy $S$ but also a  sub leading logarithmic term: 
\be
S~=~ ln~Z_{BH} ~ =~ ~\frac{2\pi r_+}{4G} ~-~\frac{3}{2}~ln~ \frac{2\pi r_+}{4G}
~+ \cdots
\ee
This is an interesting application of CS theory to 3D gravity.
In fact three dimensional CS theory also has
application in the study of black holes in four dimensional 
gravity: the boundary degrees of freedom of a black hole
in $4D$ are also described by  an $SU(2)$
Chern-Simons theory \cite{ash, kaulmaj}. This allows a
calculation of the degrees of freedom of, for example, Schwarzschild
black hole.  For large area black holes this in turn  results  in an 
expression for the entropy  which, besides a Bekenstein-Hawking area 
term, also has a logarithmic area correction with same coefficient $-3/2$
as above\cite{kaulmaj}. This suggests a universal, dimension independent,
nature of the these logarithmic corrections\cite{carlip}.

\section{\bf Topological BF theory}
There is another class of Schwarz type topological field theories 
which are known as BF theories \cite{bf1,bf2}.
These are  defined using a connection one-form A and a $(D - 2)$-form 
$B$ with values 
in the Lie algebra of a compact group $\cal G$. 
The advantage of this class of theories  is that these can be defined in 
arbitrary dimensions. While in three dimensions these theories,
like three-dimensional Chern-Simons theories, yield a description of the
topological invariants of knots and links, higher dimensional BF theories
give topological properties of higher dimensional knots (imbedded
manifolds of codimension $2$) living in these higher dimensional manifolds. 

The action for a BF theory is given by:
\be
S~=~\int_{\cM}~tr~(B\wedge F)
\ee
where $F=dA+A \wedge A$ and  now $\cM$ is a D-dimensional manifold. 
We can add a {\it cosmological term}
to this action which for  $D~=~3$ is 
\beq
S_{cos, \k } ~=~ {\frac{\k^2}{3}} \int_{\cM}~tr~(B\wedge B\wedge B) \cr
\eeq
and for $D=4$ has a form
\beq 
S_{cos, \k }~=~ {\frac{\k}{2}} \int_{\cM}~tr~(B\wedge B)
\eeq
The name cosmological term for these comes from the fact in 3D gravity,
(which can also be cast as a BF theory) this is exactly how such a 
term is written in terms of the triads.

But unlike CS theory and BF theory in three dimensions, 
it is difficult to define gauge invariant observables for
higher dimensional BF theories. However the perturbative expansion
of Wilson loop expectation values of CS theory which 
term by term corresponds to
topological invariants known as {\it Vassiliev invariants}\cite{vas}
has a generalization for BF theory in any dimensions. 

In the following, we shall briefly discuss  $D~=~3,~4$ theories.

\subsection{BF theory in D~=~3} 
The simplest case is the $BF$ theory based on $U(1)$
gauge group. 
Like in the $U(1)$ Chern-Simons theory, Abelian $BF$ theory  of
two field one-forms $A$ and $B$ also provides
a field theoretic characterization of linking and self-linking numbers of
knots. The correlator $ \la {\oint_{K_1}A}~{\oint_{K_2}B}\ra $
for two distinct knots $K_1$ and $K_2$  can easily be seen 
to be related to the  linking number of the two knots: 
$ \la {\oint_{K_1}A}~{\oint_{K_2}B}\ra $ $= i\cL(K_1,~K_2)$. 
On the other hand self-linking number of a knot is given by
 $ \la {\oint_K A}~ {\oint_K B} \ra$ $= iS\cL(K)$, where  
coincident loop correlator is defined by loop
splitting with the help of the framing knot $K_f$ as discussed earlier
in the context of $U(1)$ CS theory. 

The action for the non-Abelian BF theory including the cosmological
term $S_{BF, \k } = S + S_{cos, \k}$ is  interestingly  related to 
three-dimensional CS action\cite{bf1}:
\beq
S_{CS}(A~+~\k B)~-~S_{CS}(A~-~\k B)~~&=&~{\frac \k \pi}~S_{BF, \k },\label{grav2}\\
\left[\frac{d}{d\k}S_{CS}(A+\k B)\right]_{\k=0}~&=& {\frac 1 {2\pi}}S_{BF,
0}
\eeq

For BF theory with cosmological term, the topological operators for a 
link are constructed in terms of the operators associated with the component knots K:
~~$W[A~\pm ~\k B,K]~=~tr ~P~exp~\oint_{K} (A~\pm ~\k B)$.
Functional averages of these operators give knot/link invariants like in CS theory. 

For the theory without the cosmological term, $\k~=~0$, the topological operator
associated with a knot $K$ is:
\be
\left[\frac{d}{d\k}tr~Pexp~\oint_{K} (A~+~\k B)\right]_{\k=0}=tr\oint_K P~(exp~\int_x^y A)~B(y)~
P(exp\int_y^x A )
\ee
These are related to perturbative expansion of CS theory which in turn is related to
Vassiliev invariants\cite{vas}.
Notice that BF theories also provide a description of gravity in three dimensions;
compare Eq. \ref{grav1} and Eq. \ref{grav2}.

Next  consider the following CS functional integrals for a knot $K$:
\beq
Z_{CS}[\cM,K,k]&=&\int [\cD A]~ exp~(ikS_{CS})~ tr~P~exp\oint_KA\\  
Z_{CS}[\cM,k]&=&\int [\cD A]~ exp~(ikS_{CS})
\eeq
and the BF functional integrals:
\beq
Z_{BF,\k}[\cM,K,f]&=&\int [\cD A\cD B]~exp~\left(ifS_{BF,\k}\right)
tr~P~exp\oint_K(A+\k B)~~\\
Z_{BF,\k}[\cM,f]&=&\int [\cD A \cD B]~exp~(ifS_{BF,\k)}
\eeq
The partition functions of the BF and CS theories are related to each other 
through  following relation:
\be
Z_{CS}[\cM,k]{\overline{Z_{CS}[\cM,k]}}~=~Z_{BF,\k}[\cM,f]
\ee
where $f~=~\k k/\pi$. This relates the BF theory partition function to
Turaev-Viro invariant\cite{tur} which is the square of the CS partition function,
$|Z_{CS}|^2$. In addition the knot functional integrals in the two
theories are related as:
\be
\frac{Z_{CS}[\cM,K,k]}{Z_{CS}[\cM,k]}~=~\frac{Z_{BF,\k}[\cM,K,f]}
{Z_{BF,\k}[\cM,f]}.
\ee
This relates the knot invariants of  CS theory with those of
BF theory.

It is of interest to note that  BF theory in $D~=~3$ without 
cosmological constant provides the classic Alexander-Conway polynomial 
invariant for knots\cite{bf1}. 

\subsection{BF theory in D~=~4 } 
As mentioned earlier the BF theory  can be defined in all dimensions and
in particular in four dimensions.
The action $S_{BF, \k } = S + S_{cos, \k }$ in four dimensions has an interesting 
relation to the well known Chern-Weil form $Q_2 = \int_{\cM} tr F \wedge F$:
\beq 
Q_2(F+\k B)~-Q_2(F)~&=&~2\k S_{BF, \k }\\
\left[\frac{d}{d\k}Q_2(F+\k B)\right]_{\k =0}~&=&~2S_{BF, 0} 
\eeq
The natural geometrical setting for this case is principal bundle 
in the space of paths and loops.
The gauge fields $A$ and $B$ are collectively connections on such a principal
bundle. The observables  are generalized Wilson loops obtained 
as trace of holonomies
in the space of loops. When such observables do not involve the $B$ 
fields they are expected to be related to Donaldson invariants.  
This provides a connection of the four dimensional BF theories with
the four dimensional Witten-type topological gauge field theories\cite{witten2}.
On the other hand,  observables  involving both $A$ and $B$ fields 
are associated with embedding of 2-surfaces in the 
4-manifold as a generalization of knot theory to higher 
dimensions. Invariants characterizing these higher dimensional 
knots are obtained in theories without the
cosmological term ($\k~=~0$). These are generalizations of Vassiliev invariants of knots
in three dimensions. We refer to the literature for further references\cite{bf2}.

\vskip5mm
\noindent{\bf Acknowledgements:} The work of PR was supported by DST (India) 
grant under `` SERC FAST TRACK Scheme for Young Scientists''.


\begin{thebibliography}{9}
\bibitem{atiyah}  M. Atiyah: {\it The Geometry and Physics of Knots}, Cambridge
University Press (1989).
\bibitem{sch} A.S. Schwarz: Lett. Math. Phys {\bf 2} (1978) 247;
 
A.S. Schwarz: New topological invariants in the theory of
quantised fields, Baku International Conference (1987).
\bibitem{witten1} E. Witten: Commun. Math. Phys. {\bf 121} (1989) 351.
\bibitem{witten2} E. Witten: Commun. Math. Phys. {\bf 117} (1988) 353.  
\bibitem{kaul}R. K.  Kaul: Chern-Simons theory, knot invariants,
vertex models and three-manifold invariants, hep-th/9804122,
in {\it Frontiers of field theory, quantum gravity and
strings} (Volume 227 in Horizons in World Physics), eds. R.K. Kaul
et al, NOVA Science Publishers, New York (1999);

R. K. Kaul: Topological quantum field theories - a meeting ground for 
physicists and mathematicians, hep-th/9907119, in {\it Quantum field theory: 
a 20th century profile}, 
ed. A. N. Mitra, Indian National Science Academy, N. Delhi(2000).
\bibitem{witten3} A. Ach\'ucaro and P.K. Townsend: Phys. Letts {\bf 180}
(1986) 89;

E. Witten: Nucl. Phys. {\bf B311} (1988) 46;

S. Carlip: {\it Quantum gravity in $2+1$ dimensions},
Cambridge Monographs on Mathematical Physics (2003).

\bibitem{witten4}E. Witten: Prog. Math. {\bf 133} (1995) 637.
\bibitem{popov} A.D. Popov:  Phys. Letts. {\bf B 473} (2000) 65.
\bibitem{gauss} C.F. Gauss: Werke Vol V, G\"otingen, K\"onigliche 
Gesellschaft der 
Wissenschaften, (1833) 605, Note of January 22. 
\bibitem{maxwell} J.C. Maxwell: {\it A treatise on Electricity and Magnetism}, Oxford, Clarendon,
England, (1873). 
\bibitem{crick} F.H.C. Crick: Proc. Natl. Acad. Sciences, USA, {\bf 73} (1971) 2639.
\bibitem{deser1} S. Deser, R. Jackiw and S. Templeton: Phys. Rev. Lett. {\bf 48} (1982) 975;

S. Deser, R. Jackiw and S. Templeton: Ann. Phys. {\bf 140} (1982) 372.
\bibitem{kaulcmp} R.K.  Kaul:  Complete solution of $SU(2)$ Chern-Simons
theory, hep-th/9212129;
 
R.K. Kaul: Commun. Math. Phys. {\bf 162} (1994) 289.
\bibitem{ram}P. Ramadevi, T.R. Govindarajan and R.K. Kaul: Nucl. Phys. {\bf B402} (1993) 548.
\bibitem{birman} J.S.  Birman: {\it Braids, links and mapping class
groups}, Annals of Mathematics Studies, Princeton University Press (1975).
\bibitem{ram1} P. Ramadevi, T.R. Govindarajan, R.K. Kaul: Mod. Phys.
Letts. {\bf A9} (1994) 3205; 

P. Ramadevi, T.R. Govindarajan, R.K. Kaul: Mod. Phys. Letts. {\bf A10} 
(1995) 1635.
\bibitem{rolfsen} D. Rolfsen: {\it Knots and links}, Publish or Perish,
Berkeley (1976).

\bibitem{lic} A.D.  Wallace: Canad. J. Math. {\bf 12} (1960) 503;

W.B.R.  Lickorish: Annal of Math. {\bf 76} (1962) 531.

\bibitem{kir} R. Kirby:  Invent. Math.  {\bf 45}(1978)35;

R. Fenn and C. Rourke: Topology {\bf 18} (1979) 1.
\bibitem{kaul1}R.K. Kaul and  P. Ramadevi: Commun. Math. Phys. {\bf 217} 
(2001) 295.
\bibitem{smatrix} P.  Di Francesco, P.  Mathieu and  D.  Senechal: 
{\it Conformal Field
Theory.} Graduate Texts in Contemporary Physics, eds. J.L. Birman,
J.W. Lynn, M.P. Silverman, H.E. Stanley, Mikhail Voloshin, Springer-Verlag (1997).
\bibitem{jeff} L.C. Jeffrey: Commun. Math. Phys. {\bf 147} (1992) 563;
 
S. Kalyan Rama and S. Sen: Mod. Phys. Letts. {\bf A7} (1992) 2065.
\bibitem{ram2} P. Ramadevi and Swatee Naik:
Commun. Math. Phys. {\bf 209} (2000) 29.
\bibitem{deser2} S. Deser and R. Jackiw: Ann. Phys. {\bf 153} (1984) 405.
\bibitem{btz} M. Ba\~nados, C. Teitelboim and J. Zanelli: Phys. Rev. Letts {\bf 69} (1992) 1849.
\bibitem{lab} J.M. Labastida and A.V. Ramallo: Phys. Letts. {\bf B227} (1989)
92.
\bibitem{ours} T.R. Govindarajan, R.K. Kaul and V. Suneeta: Class. Quant. Grav. {\bf 18} (2001) 2877.
\bibitem{ash} 
L. Smolin: J. Math. Phys. {\bf 36} (1995) 6417;

C. Rovelli: Phys. Rev. Letts. {\bf 77} (1996) 3288;

A. Ashtekar, J. Baez, A. Corichi and K. Krasnov: Phys. Rev. Letts.
{\bf 80} (1998) 904.

\bibitem{kaulmaj} R.K. Kaul and P. Majumdar: Phys. Rev. Letts. {\bf 84}
(2000) 5255.
\bibitem{carlip} S. Carlip: Class. Quant. Grav. {\bf 17} (2000)
4175.
\bibitem{bf1} M. Blau and G. Thomson: Ann. Phys. {\bf 205} (1991) 130;

A S Cattaneo, P. Cotta-Ramusino, J. Fr\"ohlich and M. Martellini:
Jour. Math. Phys. {\bf 36} (1995) 6137; 

A.S. Cattaneo: Commun. Math. Phys. {\bf 189} (1997) 795.
\bibitem{bf2} A.S. Cattaneo, P. Cotta-Ramusino and  C.A. Rossi: 
Lett. Math. Phys. {\bf 51} (2000) 301; 

A.S. Cattaneo and C.A. Rossi: 
Commun. Math. Phys. {\bf 221} (2001) 591.
\bibitem{vas} V. A. Vassiliev: Cohomology of Knot Spaces, 
in {\it Theory of Singularities and Its Applications} ed. V. I. Arnold, Providence, RI: 
Amer. Math. Soc., (1990) 23.
\bibitem{tur} V.G. Turaev and O.Y. Viro: Topology {\bf 31} (1992) 865. 
\end{thebibliography}
\end{document}